\documentclass[twocolumn,twoside]{article}

\usepackage{graphicx}

\usepackage{hyperref}

\begin{document}

\thispagestyle{plain}
\markboth{\rm \uppercase{Kinematic censorship as a constraint \hspace{4cm}}}{
\hspace*{59mm} \rm \uppercase{Pavlov, Zaslavskii}}

\twocolumn[
\begin{center}
{\LARGE \bf
Kinematic censorship as a constraint on allowed\\[7pt]
scenarios of high energy particle collisions}
\vspace{12pt}

{\large \bf {Yu. V. Pavlov${}^{1,2,*}$}, \ O. B. Zaslavskii${}^{2,3,**}$}
\vspace{12pt}

{\it ${}^{1}$Institute of Problems in Mechanical Engineering,
Russian Academy of Sciences,\\
Bol'shoy pr. 61, St. Petersburg 199178, Russia;}
\vspace{4pt}

{\it ${}^{2}$N.I.\,Lobachevsky Institute of Mathematics and Mechanics,
    Kazan Federal University,\\
    18 Kremlyovskaya St., Kazan 420008, Russia;}
\vspace{4pt}

{\it ${}^{3}$Department of Physics and Technology, Kharkov V.N.\,Karazin
National University,\\
4~Svoboda Square, Kharkov 61022, Ukraine}
\vspace{4pt}

\end{center}
\vspace{5pt}
 {\bf Abstract.}
    In recent years, it was found that the energy $E_{c.m.}$ in the centre of
mass frame of two colliding particles can be unbounded near black holes.
If collision occurs exactly on the horizon, $E_{c.m.}$ is formally infinite.
However, in any physically reasonable situation this is impossible.
We collect different scenarios of such a kind and show why in every act of
collision $E_{c.m.}$ is indeed finite (although it can be as large as one likes).
The factors preventing infinite energy are diverse:
the necessity of infinite proper time, infinite tidal forces, potential barrier, etc.
This prompts us to formulate a general principle according to which the limits
in which $E_{c.m.}$ becomes infinite are never achieved.
We call this the kinematic censorship (KC).
Although by itself the validity of KC is quite natural, its application allows
one to forbid scenarios of collisions predicting infinite $E_{c.m.}$
without going into details.
The KC is valid even in the test particle approximation, so explanation of
why $E_{c.m.}$ cannot be infinite, does not require references
(common in literature) to the non-linear regime, backreaction, etc.
The KC remains valid not only for free moving particles but also if
particles experience the action of a finite force.
For an individual particle, we consider a light-like continuous limit of
a time-like trajectory in which the effective mass turns into zero.
We show that it cannot be accelerated to an infinite energy during a finite proper
time under the action of such a force. As an example, we consider dynamics
of a scalar particle interacting with a background scalar field.

\vspace{11pt}
\vspace{17pt}
]

{\centering \section{Introduction}}

\footnotetext[1]{yuri.pavlov@mail.ru}
\footnotetext[7]{zaslav@ukr.net}

Some of fundamental principles in physics have a form of prohibition. Say,
impossibility to reach the absolute zero of temperature constitutes the
third law of thermodynamics. A similar statement in black hole physics
implies that one cannot convert a nonextremal black hole into the extremal
one during a finite number of steps. The principle of cosmic censorship
states that one cannot see a singularity from the outside.
    Investigations on particle collisions revealed one more such a principle
that remained shadowed until recently.
This can be called "kinematic censorship" (KC):
the energy in the centre of mass frame of any colliding
particles cannot be infinite. In usual laboratory physics this looks quite
trivial. Indeed, in, say, flat space-time any energy gained or released in
any process is always finite. And, starting with the total finite energy,
one cannot obtain something infinite due to the energy conservation.
    The situation radically changed after findings made in~\cite{pir1,ban}.
    It was shown there that if two particles collide in the extremal Kerr
background, under certain conditions the energy $ E_{c.m.} $ in the centre
of mass becomes unbounded when a point of collision approaches
the horizon~$r_{+}$:
\begin{equation}
\lim_{r\rightarrow r_{+}}E_{c.m.}(r)\rightarrow \infty .  \label{ecm}
\end{equation}
    This was found for head-on collisions in~\cite{pir1} and
for particles moving in the same direction towards the horizon, provided
one of particles is fine-tuned, in~\cite{ban}.
    In the latter case it is called the BSW effect.

    Later on, other versions of this process were found in different contexts.
In doing so, the input was finite (particles with finite masses and energies)
but the output is formally infinite in some limiting situations.
    The fact that each time (i)~something prevents an infinite $E_{c.m.}$ and
(ii)~the nature of this "something" is completely different depending on the
concrete scenario, suggests that there exists a rather general principle
that unifies all so different particular cases. We want to point out the
following subtlety.
    The results on unbounded $E_{c.m.}$ were obtained in the test-field
approximation when self-gravitation and backreaction were neglected.
    As for sufficiently large $E_{c.m.}$ this can be no longer true,
there is temptation to ascribe formally infinite $E_{c.m.}$ to the test-field
approximation with the expectations
(confirmed in some simple models, e.g., see~\cite{shell}) that in the non-linear
regime $E_{c.m.}$ will become finite.
    However, it turns out that even
in the \textit{test-particle approximation\/} infinite $E_{c.m.}$ are
forbidden, although arbitrarily large but finite $E_{c.m.}$ are possible.
Thus the limit under discussion turns out to be unreachable in each case.

    It is worth noting that in~\cite{com} several objections were
pushed forward against the BSW effect, in particular --- impossibility
to reach the extremal state and backreaction.
    Meanwhile, it was explained~\cite{gp} that for
nonextremal black holes the BSW effect does exist, although in a somewhat
different setting. It was also shown in~\cite{rad1} for extremal black
holes and in~\cite{rad2} for nonextremal ones that account of a finite force
does not spoil the BSW effect. Therefore, the reasons why infinite $E_{c.m}$
are, nonetheless, unreachable, should have more fundamental nature and
explanations should be done just within the test-particle approximation,
without attempts to ascribe them to some factor neglected in this
approximation.

In the present work, we do not perform new concrete calculations. Instead,
we collect a number of results, already obtained earlier, and make new
qualitative generalizations based on them.

The fact that an infinite $E_{c.m.}$ cannot occur in any act of collisions,
is more or less obvious in the flat space-time.
In a curved space-time, the issue is not so trivial since formally infinite
$E_{c.m.}$ was obtained with finite initial energies of colliding particles.
    And, although from physical grounds the KC is more or less obvious,
it would be of interest to give a proof of the KC.
    However, in the present article we do not pretend for such a proof
in an arbitrary curved space-time.
    Instead, we enumerate several typical situations and trace which
conditions precisely prevent collisions with infinite $E_{c.m.}$.
    The corresponding factors turn out be rather diverse.
    We also demonstrate that the KC works as some regulator that imposes
constraint on possible scenarios and enables us to reject some of them
in advance, even without going into details.

    Another separate issue is behavior of a single particle that formally
approaches the speed of light locally in a curved background and becomes
potential source of difficulties with infinite $E_{c.m.}$.
    We suggest rather general and rigorous proof that this cannot happen
under the action of finite forces.
    This is illustrated by physically interesting example of a scalar
particle interacting with a background scalar field.

    We use the systems of units in which the speed of light $c=1$.

\vspace{4mm}
{\centering
\section{Can a massive particle be accelerated until a speed of light?}}

The question of the possibility or impossibility of infinite $E_{c.m.}$
reduces to the behavior of individual particles and their Killing energy~$E$.
If $E$ remains finite (equivalently, the velocity of a massive
particle is less than a speed of light), infinite $E_{c.m.}$ is
impossible. In the flat space-time, it is more or less obvious, that
any particle cannot reach the speed of light during a finite amount
of the proper time, provided all the components of four-acceleration
$a^{\mu }$ are finite. Meanwhile, even in such a relatively simple
case, there are some subtleties. Because of the Lorentz signature, it
would seem that although separate components of $a^{\mu }$ diverge,
the scalar $a^{2}=a_{\mu }a^{\mu }$ can, nonetheless, be finite.
However, now we argue that in the situation under discussion this is
impossible.

Statement: if a massive particle moving under the action of some
force reaches the speed of light during the finite proper time, the
absolute value of acceleration $a\rightarrow \infty $ in this point.

To the best of our knowledge, the proof of the corresponding
statement is absent from literature, so we suggest it below. We
would like to stress that the issue under discussion is not
exhausted by a simple fact that the square of the four-velocity is
equal to $-1$ even in a curved space-time and cannot turn into zero
by jump. The problem is that a time-like trajectory can approach the
light-like one continuously.
The corresponding situation was considered in the monograph~\cite{zn}
(see there Eqs. (23.2.4)--(23.2.6) and corresponding discussion).
However, the authors of~\cite{zn} restricted
themselves by a particular example whereas we consider a general
approach. Also, in the next Section we consider a particular example
how this is realized in motion of a scalar particle interacting with
the background scalar field. We demonstrate that the effective mass
vanishes in some point, so although the square of the four-velocity
remains equal to $-1$, the square of the four-momentum vanishes.

At first, let us consider the situation in the flat space-time.
Let in the Minkowski metric
\begin{equation}
ds^{2}=-dt^{2}+\eta _{ik}dx^{i}dx^{k}
\end{equation}
a particle move along the trajectory $x^{\mu }=x^{\mu }(\tau )$,
where $\tau $ is the proper time.
Then, we have
\begin{equation}
a^{\mu }=\frac{du^{\mu }}{d\tau },  \label{a}
\end{equation}
whence along a given trajectory from point 1 to point 2 one obtains
\begin{equation}
u^{\mu }=\int_{1}^{2} \! a^{\mu } d\tau ,  \label{u}
\end{equation}
where on the trajectory $a^{\mu }$ is a function of~$\tau $ only.
If all components of $a$ are finite, $u^{\mu }$ remains
finite as well, so the speed of light cannot be reached.

Now, we assume that as $\tau \rightarrow \tau _{0}$, $v\rightarrow 1$.
Here, the velocity is defined according to
$v^{2}=v_{i}v^{i} $, $v^{i}=dx^{i} / dt$.
The four-velocity has a standard form
\begin{equation}
u^{\mu } = \left( \frac{1}{\sqrt{1-v^{2}}},\frac{v^{i}}{\sqrt{1-v^{2}}}
\right) = \gamma (1,v^{i}),
\end{equation}
where $\gamma =1 / \sqrt{1-v^{2}}$ is the Lorentz factor.

For the components of $a^{\mu }$ we have $a^{0}=\dot{\gamma}$,
$a^{i}= d (\gamma v^{i}) /d\tau $.
Here, dot denotes $d / d\tau $.
After some algebra, one can obtain
\begin{equation}
a^{2}=\frac{\dot{\gamma}^{2}}{\gamma^{2}} + \gamma^{2} \frac{dv^{i}}{d\tau }
\frac{d v^{k}}{d\tau }\eta_{ik} .
\label{a2}
\end{equation}

Now, let us consider a concrete law of approaching to the speed of
light near the value $\tau =\tau_{0}$:
\begin{equation}
v^{i}=\left( v^{i}\right)_{0}-\beta ^{i}(\tau_{0}-\tau )^{n}+ \ldots ,
\end{equation}
where $\beta ^{i}$ are some constants.
Then,
\begin{equation}
v^{2}\approx 1-2\alpha (\tau_{0}-\tau )^{n},
\end{equation}
where
\begin{equation}
\alpha =\beta _{i}\left( v^{i}\right)_{0} .
\end{equation}
Then, direct evaluation of different terms in (\ref{a2}) gives us
\begin{equation}
a^{2}\approx \frac{ n^{2} }{4 (\tau -\tau_{0})^{2}}\rightarrow \infty .
\label{Eq10}
\end{equation}

Thus, not only separate components of the four-acceleration diverge
but also the scalar product does so.

In the curved space-time, the situation looks quite similar from the
physical point of view. However, it would seem that rigorous proof
becomes much more complicated.
In particular, this happens because of the appearance of the Christoffel
symbols $\Gamma_{\mu \nu}^{\rho }$ in the expression for $a^{\mu }$.
Fortunately, this is not so.
The key point here is the existence of coordinate frame in
which $\Gamma _{\mu \nu }^{\rho }=0$ along a given line.
This was shown by Fermi~\cite{fermi}.
See also textbook~\cite{rash}, Ch. VII.91.
Then, (\ref{a}) and (\ref{u}) apply with the same result.

Thus if the interval $\tau _{1}\leq \tau \leq \tau _{2}$ is finite,
$u^{\mu } $ remains finite for any finite $a$.
Otherwise, a particle
can indeed reach the speed of light but by expense that $a$ diverges
in a corresponding point. One can pass from the coordinate frame
under discussion to any other one by a finite Lorentz boost, so a
particle's speed is less than the speed of light in a new frame as
well.

The context under discussion can be considered as a rather
unexpected practical application of Fermi's finding.

To be more precise, we must make a reservation. There exists a
situation in which a particle can indeed reach a speed of light
during a finite proper time with a finite or even zero acceleration
but this is due to incompleteness and/or singular character of the frame.
For example, this happens if a particle falls in the
Schwarzschild black hole (see, say, eq. (102.7) in~\cite{LL}).
However, this velocity is measured by a static observer who becomes singular
on the horizon, so this is a consequence of impossibility of using
such a frame on the horizon and beyond. Also, the boost between the
corresponding frame and any regular frame becomes singular as well.

In the next Section, we will consider a concrete physical example in
which such a situation occurs.

\vspace{4mm}
{\centering \section{Explicit example: vanishing of the effective mass}}

In this Section, we consider a concrete physical example. We deal with a
scalar particle interacting with some background field $\psi $. Let the
action have a simple form%
\begin{equation}
S=-\int (m- q \psi )\, d \tau ,
\label{S}
\end{equation}
where $\psi$ is the background scalar field,
$q$ being the scalar charge of a particle,
$m$ is its mass.
For such a system, the exact spherically symmetric static solution
describing a black hole was found \cite{boch}--\cite{bek75}.
Its metric formally coincides with that of the extremal Reissner-Nordstr\"{o}m
black hole:
\begin{equation}
ds^{2}=-f dt^{2}+\frac{dr^{2}}{f}+r^{2}d\omega ^{2} ,
\label{met}
\end{equation}
where $f=\left( 1-\frac{r_{+}}{r}\right) ^{2}$, $r_{+}$ is the horizon radius.
Let us consider pure radial motion.
Then, one can obtain easily that%
\begin{equation}
m_{\ast }\dot{t}=\frac{E}{f}
\end{equation}
\begin{equation}
m_{\ast }\dot{r}=-P ,
\end{equation}
the effective ``mass'' $m_{\ast }\equiv m- q \psi $, the radial momentum
\begin{equation}
P=\sqrt{E^{2}-m_{\ast}^{2}f},
\label{p}
\end{equation}
the energy $E={\rm const}$.

We choose $q>0$.
Let us denote as $r_{0}$ the point in which $\psi =m/q $, so $m_{\ast }(r_{0})=0$.
It is implied that $r_{0}>r_{+}$.

We assume that $\psi $ is a smooth function of $r$ and we examine the
behavior of relevant quantities near the point $r_{0}$ where the effective
mass vanishes, $m_{\ast }(r_{0})=0$:
\begin{equation}
\psi =\psi (r_{0})-C(r-r_{0}) + \ldots ,
\end{equation}
where $C$ is some constant.
Then,
\begin{equation}
m_{\ast }\approx q C(r-r_{0})
\end{equation}

Then,
\begin{equation}
\dot{r}\approx -\frac{E}{C q (r-r_{0})} ,
\end{equation}%
for $r\geq r_{0}$, $\tau \leq \tau _{0}$
\begin{equation}
r\approx r_{0} + B\sqrt{(\tau _{0}-\tau )},
\end{equation}
\begin{equation}
B=\sqrt{\frac{2E}{C q}} .
\end{equation}%
Thus the proper time to reach $r_{0}$ (say, from $r>r_{0}$) is finite.

    For the velocity $V$ measured by a static observer, we have
$E=m_{\ast} \sqrt{f} / \sqrt{1-V^2} $, whence
\begin{equation}
V=\sqrt{1-\frac{fm_{\ast }^{2}}{E^{2}}} .
\label{v}
\end{equation}
    Near $r=r_{0}$,
\begin{equation}
V\approx 1-\frac{C^{2} q^2 f(r_{0})}{2E^2}(r-r_{0})^2 ,
\end{equation}
    so $V\rightarrow 1$.

    Thus a particle reaches the speed of light for a finite proper time but
in one point $r_{0}$ only.

One can introduce a kinematic momentum according to
$p^{\mu }=m_{\ast }u^{\mu }$.
It is instructive to note that although $u_{\mu }u^{\mu }=-1$ in
any point, including $r_{0}$, $p_{\mu }p^{\mu }=-m_{\ast }^{2}\rightarrow 0$
when $r\rightarrow r_{0}$. In this sense, we have light-like limit of a
time-like particle.
In the standard case, for a particle of a given mass,
this would lead to unbounded energy.
However, in the present case $E\,\ $ remains finite due to the fact that
$m_{\ast }\rightarrow 0$ in this limit.
As a result, we obtain a quite unusual entity --- a trajectory with
$u_{\mu} u^{\mu }=-1$ that corresponds, however,
to a massless particle in one point.

We would like to stress that these subtleties can have, in principle, further
physical consequences, say, in cosmology, where interaction of a scalar
particle with the background field can affects time evolution and conditions
of thermodynamic equilibrium~\cite{im}, \cite{ii}.

Now, we will show that price paid for such nontrivial behavior is
divergences in acceleration.
For the action~(\ref{S}), equations of motion
in this case give rise to the acceleration (see, e.g. Sec. 2 of~\cite{bek75})
\begin{equation}
a^{\alpha }=u_{;\beta }^{\alpha }u^{\beta }=
\frac{q}{m_{\ast }}[\psi^{;\alpha }+u^{\alpha }(\psi _{;\beta }u^{\beta })] ,
\label{eq}
\end{equation}
whence
\begin{equation}
a^{2}\equiv a_{\mu }a^{\mu }=
\frac{q^{2}}{m_{\ast }^{2}}h^{\mu \nu }\psi_{,\mu }\psi _{,\nu } ,
\end{equation}
\begin{equation}
h^{\mu \nu }=g^{\mu \nu }+u^{\mu }u^{\nu } ,
\end{equation}
where $u^{\mu }$ is the four-velocity.

    In general
$ \left( h^{\mu \nu }\psi _{,\mu }\psi _{,\nu }\right) _{r=r_{0}}\neq 0$
and, moreover, this quantity can diverge. This happens in the present example.
    Then, it is seen that acceleration $a^{2}\rightarrow \infty $ when
$ r\rightarrow r_{0}$.
In other words, a system as a whole is singular although all geometric
characteristics like the Kreschmann scalar are perfectly regular in
the point $r_{0}$.
It is of interest to study further properties of such ``intermediate''
systems that combine regular and singular properties.
    For the example under discussion, one can check that Eq.~(\ref{Eq10})
is reproduced exactly.

\vspace{4mm}
{\centering \section{Black vs. white holes as sources of
high energy collisions}}

    Now, we turn to concrete scenarios of high energy collisions.
    The first observations on such collisions near the horizon made
in~\cite{pir1} predict formally unbounded $E_{c.m.}$ that becomes
infinite when $r\rightarrow r_{+}$.
    However, the crucial point here consists in that
actually the corresponding scenario describes collisions of particles moving
in the opposite direction (head-on collision). This is especially clear from
Eq.~(2.57) of~\cite{pir3}, where the issue under discussion was
elaborated in detail. In turn, this implies that one of particles moves not
towards the horizon but away from it.
    This condition cannot be realized in the immediate vicinity of a black hole
horizon (see for details Sec. IV A in~\cite{fraq}) and, rather,
corresponds to a white hole.
    Collisions of particles near white holes were discussed
in~\cite{gpwhite,Zaslavskii18GC}.
    Let particle~1 appear from the inner region and collides with particle~2
that comes from infinity or some finite distance outside.
    It is essential that collision cannot happen exactly on the horizon itself.
    Otherwise, any massive particle~2 having a finite energy at infinity would
become light-like.
    Instead, one can take a particle~2 to be massless.
    However, either an observer~1 emits or absorbs a photon of finite frequency
or the frequency of particle~2 must be taken infinite from the very beginning
that deprives the scenario of physical meaning~\cite{along}.
    Thus, nature of the white horizon protects the energy $E_{c.m.}$ from being
infinite.

\vspace{4mm}
{\centering \section{Extremal black holes}}

    Let us consider the metric
    \begin{equation}
ds^{2}=-N^{2}dt^{2}+g_{\phi }(d\phi -\omega dt)^{2}+\frac{dr^{2}}{A}
+g_{\theta }d\theta ^{2}
\end{equation}%
in which the coefficients do not depend on $t$ and $\phi $. We assume that
the function $\eta \equiv \sqrt{A}/N$ is finite on the horizon like
in the Kerr-Newman metric. By definition of an extremal black hole, near the
horizon~$r_{+}$,
    \begin{equation}
N\sim r-r_{+} .  \label{en}
\end{equation}

For equatorial motion, it follows from the geodesic equations that the
proper time $\tau $ required for a particle to travel from $r_{0}$
to $ r<r_{0}$ towards the horizon equals
    \begin{equation}
\tau =\int_{r}^{r_{0}}\frac{m \eta\, dr}{\sqrt{(E-\omega L)^{2}-N^{2}(m^{2}+
\frac{L^{2}}{g_{\phi }})}} .
\end{equation}
    Here, $E$ is the particle energy, $L$ being its angular momentum, $m$ mass.

    The essential feature of the Ba\~{n}ados-Silk-West effect is
that one of particles has $ E-\omega_{H} L=0$,
where $\omega _{H}$ is the value of $\omega $ on the horizon.
    Such a particle is called critical.
    For small $N$ we have $E-\omega L=O(N)$.
    Taking into account (\ref{en}), we see that $\tau \sim \left\vert \ln
(r-r_{+})\right\vert $ diverges~\cite{gp,prd,ted}.
    It means that a fine-tuned particle never reaches the horizon, so
collision never occurs exactly on the horizon. It can happen close to it.
    Then, $E_{c.m.}$ is large (even unbounded) but finite in each concrete
collision.
    The same situation happens for collision of radially moving particles in
the background of the Reissner-Nordstr\"{o}m metric~\cite{jl}.

\vspace{4mm}
{\centering \section{Nonextremal black holes, collisions outside}}

\begin{figure}[th]
\centering
\includegraphics[width=0.46\textwidth]{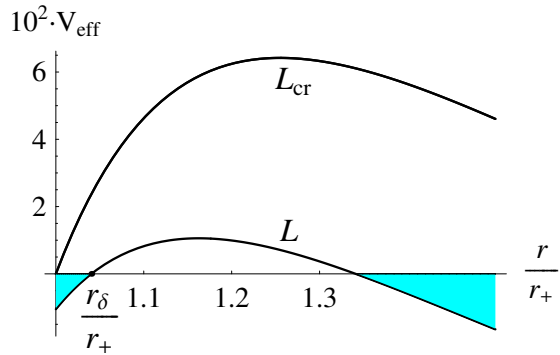}
\caption{The effective potential for motion in the equatorial plane of
the Kerr black hole with $a/M = 0.95$ and
a critical particle with $L_{\rm cr} \approx 2.76 M m $.
Allowed zones for a particle with $L = 2.5 M m $ are colored.}
\label{Fig1}
\end{figure}

    Now, there are no trajectories with infinite $\tau $.
    However, another difficulty comes into play.
    The allowed region of motion is characterized by $V_{\rm eff} \le 0 $
(see Fig.~\ref{Fig1}).
    Here, the effective potential $V_{\rm eff}$ is defined according to
(see~\cite{potent} for details)
    \begin{equation}
\left(\frac{dr}{d\tau }\right)^{2}+V_{\rm eff}=0.
\end{equation}
    The critical particle cannot approach the horizon at all because of
the potential barrier.
    One can let parameters of a particle to differ slightly from those
of the critical one.
    Then, such a near-critical particle can approach the horizon and collide
there with some another one. In such a case $E_{c.m.}\sim 1/\sqrt{\delta }$
where $\delta $ is the parameter that controls the deviation
from the exact critical relationship~\cite{gp}.
    Thus, $E_{c.m.}$ can be made as large as one likes but not infinite.
And, collision of interest can occur within a narrow strip near the horizon
only~\cite{gp,prd}:
    \begin{equation}
0 \le N \le N_{0} ,
\end{equation}
where $N_{0}\sim \delta $.
For the Kerr metric (see Fig.~\ref{Fig1}) $ r_+ \le r \le r_\delta $.
When $\delta \rightarrow 0$, the region with high
energy outcome degenerates into the point, so collision between two
particles becomes impossible.

\vspace{4mm}
{\centering \section{Nonextremal black holes, collisions inside}}

    Now, both potential barrier and infinite proper time are not encountered
in the problem.
    Formally, when a particle approaches the inner horizon $r_{-}$,
    \begin{equation}
\lim_{r\rightarrow r_{-}}E_{c.m.}(r)\rightarrow \infty .
\label{7}
\end{equation}

    The example from the present Section is especially instructive since it
demonstrates the predictive power of the principle under discussion.
    Several years ago, an intensive discussion developed in literature
concerning the possibility of the analogue of the BSW effect inside
a black hole.
    Firstly, it was claimed in~\cite{Lake10}
that such an analogue does exist.
    Later on, the author himself refuted this result~\cite{lake2},
independently this was done in~\cite{gp-astro}.
    More weak version of this effect revived in~\cite{inner}.
    Another examples of the same kind (with predictions of infinite $E_{c.m.}$)
were suggested in~\cite{gao} for the nonextremal Kerr metric and in~\cite{zhong}
for the cosmological horizon.
    These results are incorrect, as is explained in~\cite{inner}.
        The reason consists in that relevant trajectories
(that otherwise would have given infinite $E_{c.m.}$) do not intersect on
the horizon.

    Meanwhile, such scenarios can be rejected at once only due to
contradiction to our KC.
    This principle is able to reject the scenario described
in~\cite{Lake10,gao,zhong} even without elucidating these details
(which can be found in aforementioned references).
    Indeed, (i) the result~(\ref{7}) is \textit{always\/} formally valid  for
\textit{any two\/} particles,
(ii) Eq.~(\ref{7}) implies that an infinite $E_{c.m.}$ is reachable in a separate
act of collision.
    But this is what our principle forbids!

\vspace{4mm}
{\centering \section{Scalar field and infinite acceleration}}

Let a black hole be surrounded by a scalar field and one of colliding
particles is minimally coupled to this field.
The corresponding action describing interaction of a scalar particle
with the background scalar field is described by Eq.~(\ref{S}).
However, now we change the sign of the particle's scalar charge~$q$.
Then, there are no divergences for acceleration outside the horizon.
But, instead, another phenomenon comes into play.
It is connected with the immediate vicinity of the horizon.
Let the scalar field diverges near the horizon like $\varphi \sim N^{-\beta} $.
If $\beta <1$, the proper time of traveling to to the horizon is
indeed finite~\cite{sc}. Meanwhile, if one calculates the absolute value of
the four-acceleration experiences by a particle under action of the scalar
field, $a\sim N^{2(\beta -1)}\rightarrow \infty $ near the horizon~\cite{sc}.
The gradient of $a$ diverges as well.
Any particle having small but nonzero size will be teared to pieces,
so collision on the horizon becomes impossible.

\vspace{4mm}
{\centering \section{From black holes to naked singularities}}

Up to now we considered collisions in the background of black holes that is
the main subject of our work since it is this case when KC is nontrivial.
Meanwhile, we would like to make a short comment how KC reveals itself for
naked singularities.
Although this case is much more simple, it is quite instructive.

It was noticed in \cite{pat} that naked singularities can serve as
accelerators in the two-step scenario.
    One of particles moves from infinity, reflects from the potential barrier
and collides with the second particle falling from infinity in the point
where $N$ is very small, i.e. on the verge of forming the horizon
(but it does not form).
Then, $E_{c.m.}^{2}\sim N_{c}^{-2}$, where $N_{c}$ is the value of $N$ in
the point of collision.
By its very meaning, $N_{c}$ cannot reach the value $N_{c}=0$ since
this would mean formation of the horizon, so KC holds true.

\vspace{4mm}
{\centering \section{Conclusion}}

    We considered several completely different examples in which it seemed
to be "obvious" that $E_{c.m.} $ could be infinite.
    However, we saw that in each of the examples, some "hidden" factor
reveals itself which acts to prevent $E_{c.m.}$ from being infinite.
    The nature of these factors is quite different and depends strongly
on the situation. This indeed points to the validity of some fundamental
underlying principle called by us "principle of kinematic censorship".
    From a more practical point of view, this principle, by itself,
cannot suggest some new scenarios of high energy collisions.
    Rather, it can serve as some constraint that enables us to separate
forbidden scenarios from physically possible ones and reveal the factors
(sometimes hidden) that make $E_{c.m.}$ impossible.

    Thus we formulated some new principle, suggested arguments in its
favour and checked it on concrete examples.
    There are two qualitative lessons following from our analysis:
(i)~there is a crucial difference between infinite and
unbounded $E_{c.m.} $ (finite in each act of collision),
(ii)~the impossibility of infinite $E_{c.m.} $ is inherent to any scenario
in the \textit{test particle approximation}.
    There is no need to refer to some not quite understandable factors which
will be taken into account in future investigations of non-linear regimes!

    In addition to results concerning the properties of collisions,
we gave proof for a curved background that under the action of a finite force,
a trajectory of a massive particle cannot become light-like.
    In other words, a massive particle cannot turn into a massless one.
    This can be of some interest on its own in other contexts~\cite{Mann}.

\vspace{4mm}
{\bf Acknowledgments.}
This work was supported by the Russian Government Program of Competitive
Growth of Kazan Federal University. The work of Yu.\,P. was supported also by
the Russian Foundation for Basic Research, grant No. 18-02-00461-a.
O.\,Z. is grateful to Serguei Krasnikov and Alexey Toporensky for discussion
of Sec.~2 and to Kirill Bronnikov for interest to materials of Sec.~3.

\vspace{4mm}

\end{document}